# Large-scale integrated reconfigurable orbital angular momentum mode multiplexer

Ning Zhang[1*], Mirco Scaffardi[2], Charalambos Klitis[1], Muhammad N. Malik[2,3], Veronica Toccafondo[2], Francesco Fresi[3], Jiangbo Zhu[4], Xinlun Cai[5], Siyuan Yu[4,5], Antonella Bogoni[2,3], and Marc Sorel[1,3*]

[1]*School of Engineering, University of Glasgow, Rankine Building, Oakfield Avenue, Glasgow G12 8LT, UK.*
[2]*CNIT, Via Moruzzi 1, 56124 Pisa, Italy.2*
[3]*Scuola Superiore Sant'Anna, Via Moruzzi 1, 56124 Pisa, Italy.*
[4]*Photonics Research Group, School of Computer Science, Electrical and Electronic Engineering, and Engineering Maths, University of Bristol, Woodland Road, Bristol BS8 1UB, UK.*
[5]*State Key Laboratory of Optoelectronic Materials and Technologies and School of Physics and Engineering, Sun Yatsen University, Guangzhou 510275, China.*
[*]*Correspondence and requests for materials should be addressed to N. Zhang (email: Ning.Zhang@glasgow.ac.uk) or M. Sorel (email: Marc.Sorel@glasgow.ac.uk).*

Recent experiments in orbital angular momentum multiplexing have demonstrated its potential for improving the link capacity in optical interconnection networks. Meanwhile, compact photonic integrated orbital angular momentum (de-)multiplexing devices are needed to address requirements such as high scalability, fast configurability, low cost and low power consumption. Here we report on a large-scale integrated tunable orbital angular momentum mode multiplexer, which supports up to 20 multiplexed orbital angular momentum modes over 16 wavelength channels with 30 GHz channel spacing. A testbed of nine multiplexed OAM beams encoded with 28 Gbaud 16-quadrature amplitude modulation signal is demonstrated, achieving a 1.008 Tbit s$^{-1}$ aggregated rate with low penalty and sub-microsecond reconfiguration rate of the orbital angular momentum mode. This device offers an effective solution for replacing bulky diffractive optical elements, paving the way for orbital angular momentum multiplexing in optical interconnection networks.

Driven by the growth of data in large information technology (IT) infrastructures, the demand for optical interconnection networks with large bandwidth is rapidly increasing. Such demand presents severe challenging technological issues in terms of the scalability and performance of the current electrical interconnections[1]. Though in the past, the link capacity has been largely increased by wavelength division multiplexing (WDM), deploying new forms of data channel multiplexing has become essential to meet the ever-growing bandwidth requirements[2,3]. A common option to enhance the total capacity of optical interconnections is the use of orthogonal spatially overlapping and co-propagating spatial modes, known as Mode Division Multiplexing (MDM). One type of MDM that has recently attracted significant interest is orbital angular momentum (OAM) multiplexing[4,5].

OAM beams are characterized by an additional azimuthal phase term $\exp(jl\varphi)$, where $l$ an integer with either positive or negative sign, is the topological charge of the beam and represents the quanta of the OAM state[6]. Owing to the fact that OAM beams with different topological charges are mutually orthogonal[7], OAM beams can provide multiple independent data channels in the same spatial channel. Furthermore, the OAM of an electromagnetic (EM) wave is independent of other domains, so that OAM multiplexing is compatible with other multiplexing domains, such as WDM and polarization division multiplexing (PDM)[8].

The first experimental demonstration of optical OAM multiplexing in the C band was reported in 2010[9], where two 10 Gbit s$^{-1}$ multiplexed OAM channels were transmitted error-freely in free space. Later, a number of OAM multiplexing systems were demonstrated with higher transmission capacity and spectrum efficiency[10-12]. Meanwhile, multiplexed OAM beams were successfully transmitted in a vortex fibre over a length of 1.1 km, which set off the trend of large capacity and low crosstalk OAM fibre research[13-17]. Though OAM multiplexing technology has proved to significantly enhance the interconnection capacity, the bulk spatial light modulators (SLM) and spiral phase plates (SPP) currently used to manipulate the beams cannot meet the typical requirements of optical interconnection networks such as small form factor, low-cost, low-power consumption and fast configurability[2]. Because photonic integration offers an ideal technology to meet these requirements, a number of devices such as integrated circular grating couplers[18] and micro-rings with angular gratings[19-21] were reported as more compact alternatives for OAM generation. Additionally, two channel free-space OAM multiplexing based on a photonic integrated circuit was demonstrated by transmitting 10 Gbit s$^{-1}$ binary phase shifted keying (BPSK) data to achieve a 20 Gbit s$^{-1}$ aggregated rate[22,23]. However, these integrated devices were complicated to reconfigure and their OAM multiplexing capability was limited to a few modes, and hence they do not offer a satisfying solution for optical interconnection systems.

In this paper, we propose a large-scale integrated OAM multiplexer based on concentric Ω-shaped silicon waveguides. This technologically simple device allows to multiplex WDM channels on OAM beams, which can be independently tuned over 10 OAM modes. In the proof-of-concept demonstration, we show a compact ten-layer OAM multiplexer which can support up to 20 multiplexed OAM modes over 16 WDM channels with 30 GHz channel spacing. Combining such large figures in terms of WDM channels, OAM layers and tunability in a single integrated device offers a substantial advance with respect to the state-of-the-art. Moreover, this device is fully packaged and successfully evaluated in a testbed where nine multiplexed OAM beams are encoded with 28 Gbaud 16-quadrature amplitude modulation (QAM), demonstrating a 1.008 Tbit s-1 aggregated rate. The demodulated OAM beams show an optical signal to noise ratio (OSNR) penalty <2.8 dB and a power consumption <0.1 mW (Gb s$^{-1}$)$^{-1}$. Furthermore, the measured sub-microsecond reconfiguration rate between OAM modes makes this device a potential candidate for OAM switching in optical interconnection networks.

## Results

**Integrated tunable OAM multiplexer.** The working principle of the proposed OAM multiplexer is illustrated by the schematic of Figure 1a. The emitters are formed by two input/output bus waveguides connected to a quasi-circular geometry, which resembles an *omega* (Ω) shape. The aperture (α) in the circular geometry enables the integration of multiple coaxial emitters without any crossings between the bus waveguides. Because the aperture causes a non-uniform near-field intensity distribution in the azimuthal direction, the maximum angle α of the aperture is kept below 60° to ensure a satisfactorily modal purity and a low crosstalk between the modes. The access waveguides are connected to the Ω-shaped geometry through an Euler bend to minimize the modal mismatch[24]. Second-order Bragg gratings are patterned on the inner sidewall of the waveguides to scatter the guided optical modes to vertically radiated OAM modes, which consist of two components – a right-hand circularly polarized (RHCP) beam and a left-hand circularly polarized (LHCP) beam[19]. The grating profile of each waveguide layer is a narrow square, the design of which was optimised using a coupled-mode theory (CMT)-based model[25,26] to enhance one of the circular polarized components through weakening the in-plane optical back-reflection inside the Ω-shaped silicon waveguide[27]. Meanwhile, the grating profiles of each Ω-shaped waveguide emitter are designed to keep the vertical emission efficiency of each emitter uniform, i.e. the κ·L product between the grating strength κ and the length of the emitter L is constant for all the emitters. Metallic heaters are defined in close proximity to each Ω-shaped waveguide to thermally tune the radiated OAM mode independently, which allows different Ω-shaped waveguide to emit beams carrying different OAM states. The topological charge ($l_i$) carried by the OAM beam emitted from the $i^{th}$ Ω-shaped waveguide ($Ω_i$) when light is coupled in from one side can be expressed as:

$$l_i = \frac{(n_{e_i} + \Delta n_{e_i})L_i}{\lambda_{in}} - q_i \quad (1)$$

where $n_{ei}$ is the effective refractive index of the $i^{th}$ waveguide, $\Delta n_{ei}$ is the refractive index change due to the thermal power applied on the $i^{th}$ metallic heater, $\lambda_{in}$ is the input wavelength, $L_i$ and $q_i$ are the length and number of grating of the $i^{th}$ Ω-shaped waveguide, respectively. If the light is coupled into the waveguide from the opposite side the topological charge of the emitted OAM beam becomes -$l_i$.

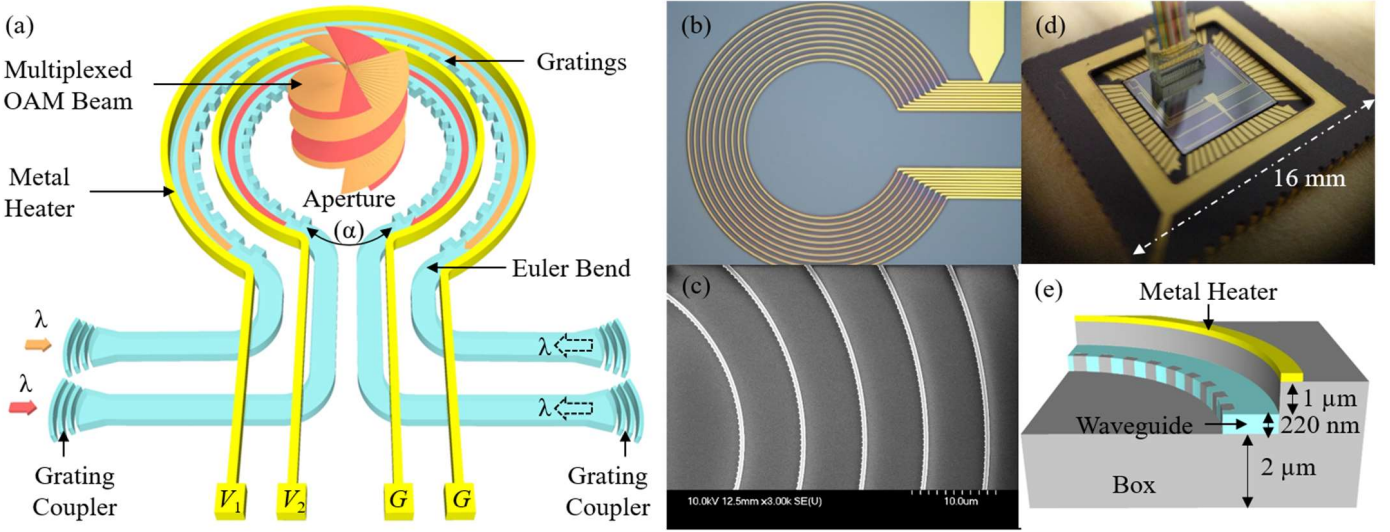

**Figure 1.** Illustration of the integrated reconfigurable OAM multiplexer. (a) 3D schematic of a two-concentric Ω-shaped OAM multiplexer. The vertically emitted beams are two co-axially multiplexed OAM beams. The optical signals are coupled into the waveguide emitters through grating couplers defined on both sides of the bus waveguides. $V_1$ and $V_2$ represent the voltages applied to the metallic heaters to tune the emitted OAM state, while *G* represents the ground connection. (b) Microscope picture and (c) SEM picture of a fabricated OAM multiplexer consisting of ten-concentric Ω-shaped emitters. (d) Picture of a fully packaged OAM multiplexer with the electrical connections and an optical fibre array. (e) Cross section of the waveguide layer structure showing the silicon core layer with the sidewall gratings, the silica cladding and the metallic heater for the tuning.

Based on the working principle described above, we designed a device with ten-concentric Ω-shaped waveguides to allow 20 OAM channels to be multiplexed when light is coupled into the waveguides from both sides simultaneously. Figure 1b and 1c show an optical microscope and a scanning electron microscope (SEM) image of the fabricated OAM multiplexer with ten layers. The radii of the Ω-shaped waveguides range from 69.54 μm to 118.95 μm with a step of 5.49 μm while the apertures of the Ω-shaped waveguides range from 45° to 50°. These geometrical parameters are the results of a compromise between small device size and low optical and thermal cross coupling between the adjacent Ω-shaped waveguides and access waveguides[28]. Figure 1d shows the silicon chip packaged into a 16 mm x 16 mm square ceramic chip carrier with bonded electric wires to tune the OAM multiplexer and a pigtailed fiber array to couple the light into the device. The device is fabricated on a silicon-on-insulator (SOI) wafer, with the waveguide cross-section structure depicted in Figure 1e.

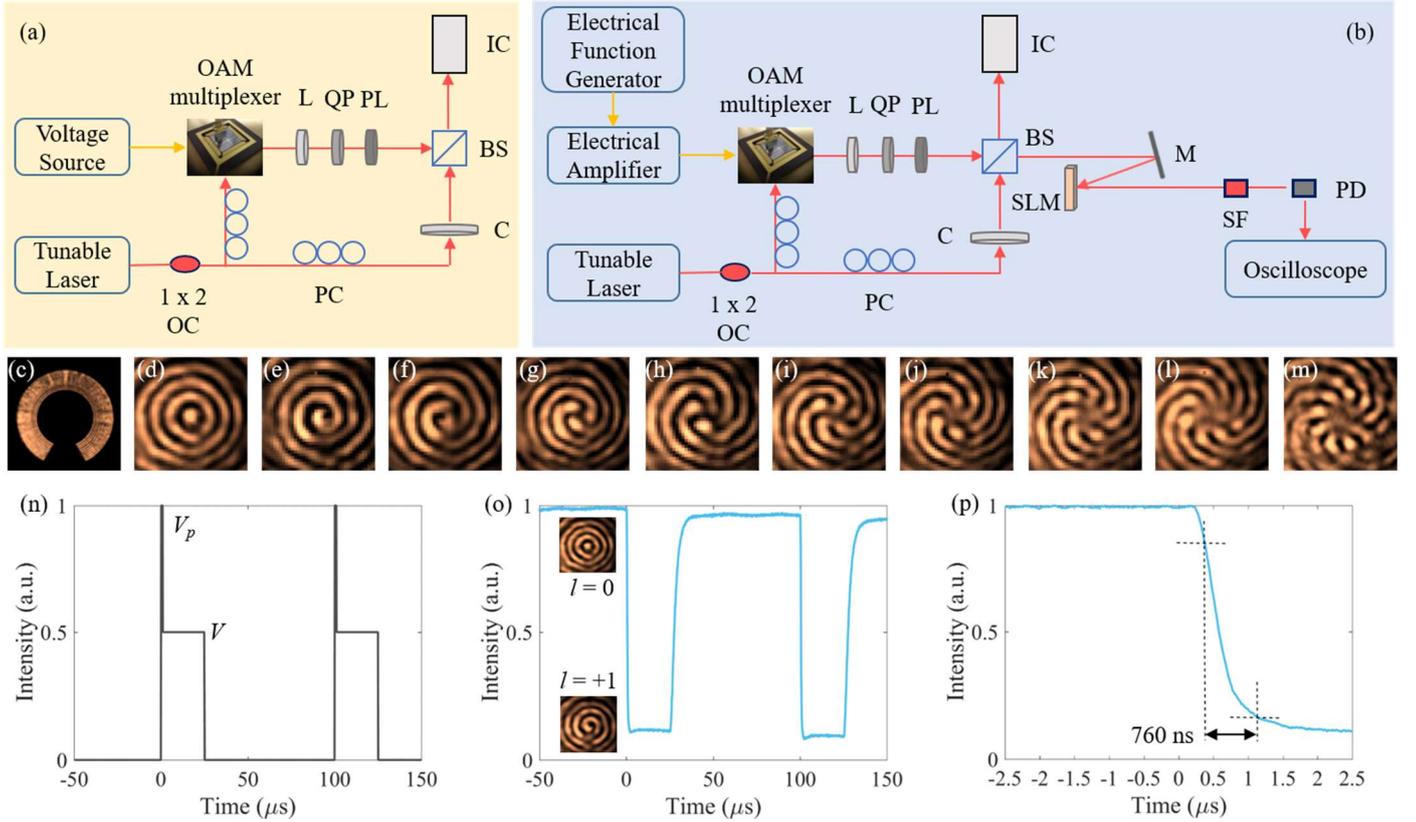

**Figure 2**. Schematics of the setups used for the characterisation of the integrated reconfigurable OAM multiplexer. (a) Setup for static characterisation. OC: optical coupler. PC: polarization controller. L: lens. QP: quarter wave plate. PL: polarizer. C: collimator. BS: beam splitter. IC: infrared camera. (b) Setup for dynamic characterisation. M: mirror. SF: Single mode fibre. PD: power detector. (c) Near-field intensity of the Ω-shaped OAM multiplexer with ten-concentric emitters. (d-m) Interferograms of the OAM beams emitted from each Ω-shaped waveguide, from the smallest (d) to the largest (m) device. The input wavelength is 1552.75 nm and the OAM orders are $l$ = 0, -1, -2, -3, -4, -5, -6, -7, -8, -9. (n) Normalised electrical driving signal. (o) Measured optical signal for switching between $l$ = 0 and $l$ = +1 on the OAM beam emitted by $\Omega_1$. (p) Zoom-in of the falling edge of the measured optical signal for switching between OAM order $l$ = 0 and $l$ = +1. Although results here are only presented for $\Omega_1$, very similar switching speeds were recorded on the other emitters.

**Device characterization.** The setups for the static and dynamic characterizations of the integrated tunable OAM multiplexer are shown in Figure 2a and 2b. A lens (L) in front of the multiplexer collimates the beam, a quarter wave plate (QP) and a linear polarizer (LP) select the dominant circularly polarized component, an infrared camera (IC) monitors the near-field and interferograms patterns and an oscilloscope after the photodetector (PD) measures the tuning response after the OAM beam is demodulated by the SLM.

Figure 2c is a picture of the observed near-field intensity profile when light (λ=1552.75 nm) is simultaneously coupled into the ten-concentric Ω-shaped waveguides. The emission efficiency of each Ω-shaped waveguide, which is defined as the ratio between the power collected after the collimated lens, $P_{out}$, and the power coupled into the grating coupler, $P_{in}$, is around 20%. Figure 2d to 2m represent the spiral interferograms produced by the OAM beams radiated from each Ω-shaped waveguide. In this case, light is coupled into the waveguides one at a time and the emitters are tuned to emit OAM beams with topological charge $l$ ranging from 0 to -9.

To improve the response time of the switching between OAM beams with different topological charge, the step-like electrical signal applied to the heaters is preceded by a short excitation pulse as shown in Figure 2n[29]. The peak voltage of the pulse is twice that of the voltage needed to tune the emitted beam over one OAM state. The total tuning range is over 10 OAM states with an efficiency of 26 mW per state. The time trace of the switching between OAM modes with charge $l = 0$ and $l = +1$ together with their interferogram patterns are shown in Figure 2o. The falling edge in Figure 2o represents the switching from the OAM state with $l = 0$ to the adjacent OAM state with $l = +1$, while the rising edge represents the opposite process. The benefit of adding an excitation pulse to the driving signal can be clearly appreciated by inspecting the time trace of Figure 2o: the fall time (from 90% to 10% of the maximum amplitude) is measured as 760 ns, while the rising edge, which is not preceded by any excitation pulse, shows a recovery time of around 10 μs. The tuning time can be further improved to values below 100ns by using an electrical signal with a stronger excitation pulse[29].

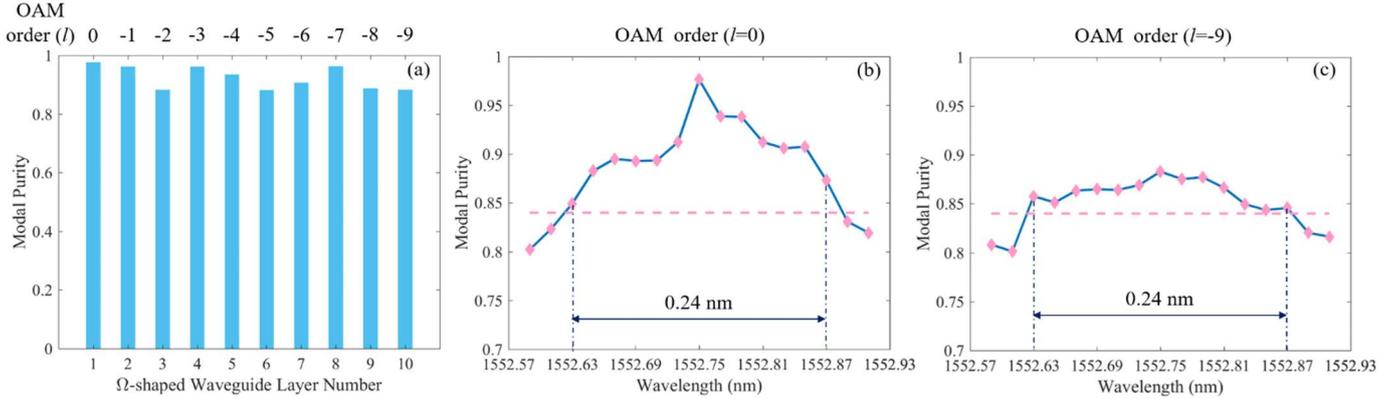

**Figure 3**. The modal purity and optical linewidth measurements of the integrated OAM multiplexer. (a) The modal purity measurement of the OAM mode emitted from each Ω-shaped waveguide. The input wavelength is 1552.75 nm and the OAM orders range from $l = 0$ to $l = -9$. (b) Optical linewidth measurement of $\Omega_1$ and (c) $\Omega_9$.

**OAM modal purity and transmission bandwidth.** The modal purity of the emitted OAM signal is an important device characteristic as it strongly impacts on the signal cross talk. In our measurements, the OAM modal purity is defined by $P_l/P_{sum}$, where $P_l$ is the power of the intended OAM state with order $l$ and $P_{sum}$ is the power of all the OAM states carried by the emitted beam ranging from $l - 10$ to $l + 10$ to achieve statistic accuracy. The measurement of the OAM modal purity was carried out with the setup depicted in Figure 2b, where each OAM state is individually evaluated by measuring its power after demodulation with the SLM[20]. Each waveguide layer is tuned to emit a different topological charge from $l = 0$ to $l = -9$ (see Figure 2 d-m) and the modal purities of the OAM modes emitted from the ten-concentric Ω-shaped waveguide range from 88% to 97%, which is comparable to the performance of previously reported geometries such as the ring-waveguide OAM beam emitter[20].

Furthermore, the modal purity is measured as a function of the input wavelength to assess the total transmission bandwidth each Ω-shaped OAM emitter can support. Figure 3b and 3c show the measurement results of $\Omega_1$ and $\Omega_9$, both of which indicating that the purity of the OAM order $l$ worsens as the wavelength detuning increases. Considering a minimum crosstalk to neighboring OAM channels of 10 dB (i.e. 84%), the total bandwidth is around 0.24 nm, a value that can support signal modulations up to 30 GHz.

**Transmission performance.** The transmission performance of the integrated tunable OAM multiplexer was measured by tuning both the input signal wavelength and the order of the emitted OAM mode. Specifically, the OAM order $l$ and the wavelength $\lambda$ were varied on each emitter over an interval $l_1 - l_2 = \Delta l = 10$ and $\lambda_2 - \lambda_1 = \Delta\lambda = 3.84$ nm, respectively. The chosen $\Delta l$ and $\Delta\lambda$ values correspond to the maximum thermal tuning range of the metallic heaters and the spectral occupancy of 16 WDM channels with a 30GHz channel spacing.

The experimental setup is shown in Figure 4b. The OAM beam is emitted from a narrow waveguide that generates a Bessel beam-like far-field intensity distribution[19]. After being collimated by a lens, the OAM beam emitted from the Ω-shaped waveguide propagates diffraction-free for a distance of D·f/d$_Ω$, where D and f are the diameter and focal length of the collimator and d$_Ω$ is the diameter of the Ω-shaped device[30,31].

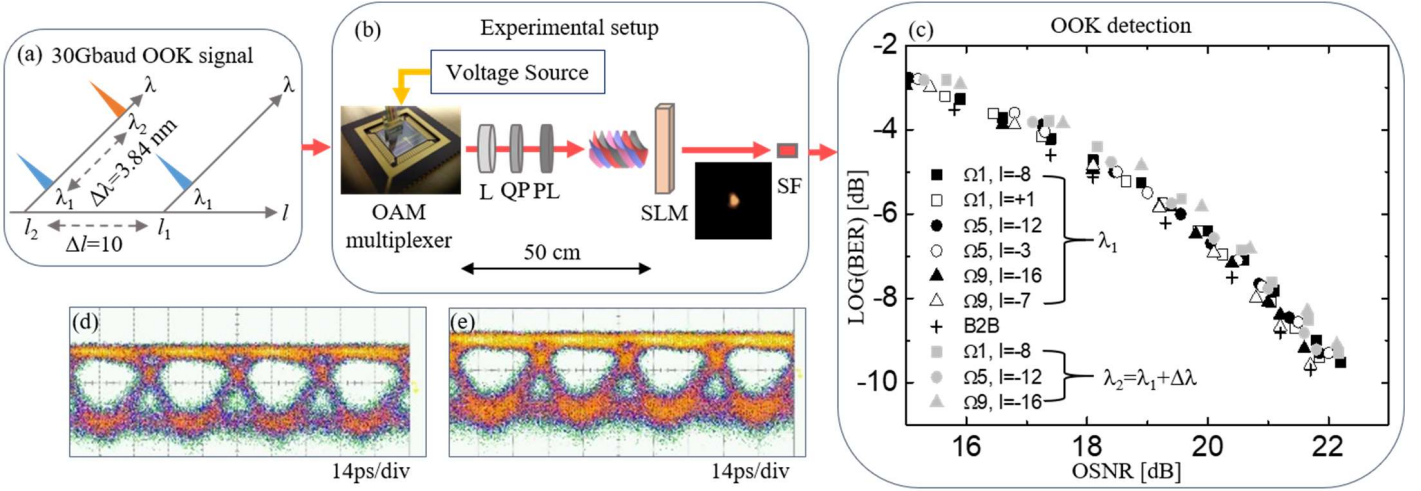

**Figure 4**. Transmission performances of each Ω-shaped OAM emitter (a) The performances are measured by tuning both the wavelength of the input signal (from λ$_1$ = 1557.55 nm to λ$_2$ = λ$_1$ + Δλ, with Δλ = 3.84nm) and the order of the emitted OAM mode (from $l_i$ to $l_i$ + Δ$l$, with Δ$l$ = 10). (b) Experimental setup where the symbols used to indicate the various components are the same as those of Figure 2. The propagation distance between the multiplexer and the SLM is 50 cm. (c) Transmission performance of the integrated OAM multiplexer. The BER versus the OSNR is reported for waveguides Ω$_1$, Ω$_5$ and Ω$_9$. (d) Eye diagram of the signal coupled to the OAM multiplexer. (e) Eye diagram of the signal converted to OAM of order $l$ = -12 by Ω$_5$ and converted back to a Gaussian beam by the SLM.

The Ω-shaped waveguides are fed one at a time with a 30 Gbaud on-off keying (OOK) pseudo-random bit sequence (PRBS) $2^7$ – 1. Figure 4c shows the bit error rate (BER) versus the OSNR for Ω$_1$, Ω$_5$ and Ω$_9$. For each Ω-shaped waveguide, the BER curves are reported for two emitted OAM modes ($l_i$ and $l_i$ + Δ$l$) at a fixed wavelength λ$_1$ = 1557.55 nm, and for OAM mode $l_i$ at a wavelength λ$_2$ = λ$_1$ + Δλ. In all the examined cases, the OSNR penalty with respect to the back-to-back (B2B) at BER $10^{-9}$ is <1 dB, which demonstrates that the integrated OAM multiplexer has a uniform performance over a set of 10 OAM modes and a range of wavelengths corresponding to 16 WDM channels with a 30 GHz channel spacing.

The eye diagrams of the signal at the input of the OAM multiplexer and at the output of the OAM demodulation (SLM) are shown in Figure 4d and 4e respectively. The output eye is open and not distorted by the OAM modulation and demodulation process.

The integrated OAM multiplexer was also tested in an experiment where the Ω-shaped waveguides were fed with the same 28 Gbaud 16-QAM PRBS $2^{15}$ - 1 at 1557.55 nm, whose constellation is shown in Figure 5a. The waveguides are thermally tuned in order to emit beams carrying different OAMs, in such a way that if $l_i$ is the OAM emitted by Ω$_i$, $l_i$ + 1 is emitted by Ω$_i$ - 1 and $l_i$ - 1 is emitted by Ω$_i$ + 1. This allows to evaluate the performance in the worst possible crosstalk condition when contiguous waveguides emit contiguous OAM modes. The experimental setup shown in Figure 5b is the same of that described in Figure 4b with the addition of a 1 × 10 optical coupler. Figure 5c reports the BER curves as a function of the OSNR for the signal emitted by each Ω-shaped waveguide. At the BER forward error correction (FEC) limit of 2.4 x $10^{-2}$ (corresponding to log(BER) = -1.62), the penalty with respect to the B2B is <2.8 dB. Figure 5d shows the constellation of the signal converted to an OAM of order $l$ = -12 emitted by Ω$_5$ and converted back to a Gaussian beam by the SLM. Although the crosstalk between different OAM states reduces the space among the constellation with respect to the B2B, the measured performance makes this integrated OAM multiplexer suitable for medium-short point-to-point links for metro and data-center applications.

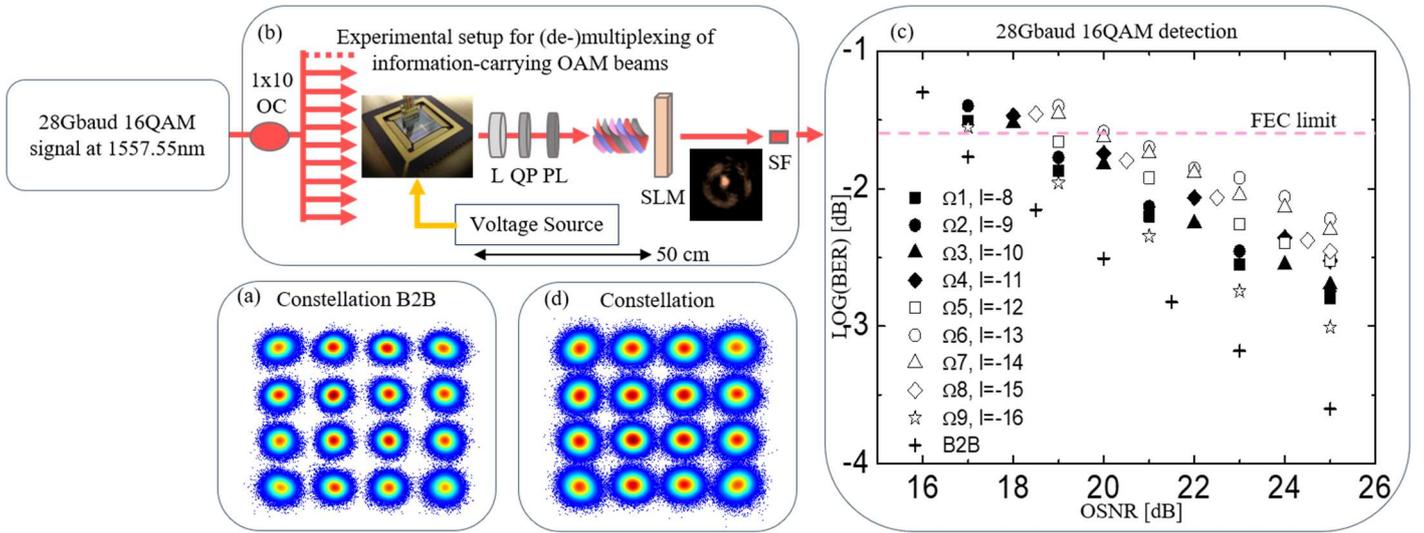

**Figure 5**. Transmission performances of multiplexed OAM beams. (a) Constellation of the input 28 Gbaud 16-QAM signal at 1557.55 nm. (b) Experimental setup. The insert shows the pattern of the transmitted OAM beam after demodulation. (c) BER versus OSNR for all the nine Ω-shaped waveguides fed with the same 28 Gbaud 16-QAM signal at 1557.55 nm. Contiguous waveguides are set to emit consecutive OAM modes. (d) Constellation of the signal first converted to OAM of order $l = -12$ emitted by $\Omega_5$ and then converted back to a Gaussian beam by the SLM.

## Discussion

To sum up, we have demonstrated an innovative integrated OAM multiplexer that shows full and fast reconfigurability of 10 multiplexed OAM beams over 16 WDM channels. The device is fully packaged and tested in a 28 Gbaud 16-QAM optical coherent system. The observed <2.8 dB OSNR penalty and <0.1 mW (Gb s$^{-1}$)$^{-1}$ power consumption confirm the potential of this technology for optical interconnection networks. The device can be further improved in terms of modal crosstalk and OSNR penalty by decreasing the aperture angle of the Ω-shaped waveguide. In addition, the grating profiles can be designed with an exponential increase along the waveguide propagation direction[26] to keep the vertical emission more uniform along the azimuthal dimension. More complex grating designs can also be engineered to achieve, for example, the simultaneous emission of OAM modes with different orders from the same emitter[21].

A further possible application of the proposed integrated OAM multiplexer is in a cascade configuration with an OAM sorter[32,33] to build a two-layer switch architecture exploiting both the OAM and the wavelength as switch domains as demonstrated in a recently published paper[34]. This architecture enables the implementation of high-throughput optical switches in a compact fashion, since it employs a single integrated device for OAM modulation and multiplexing and a single small-form factor device[32,33] for OAM demultiplexing and demodulation.

**ACKNOWLEDGMENTS**

This work has been funded by the EU Horizon 2020 Project ROAM (Call ID: H2020-ICT-2014-1; topic: ICT-06-2014; funding scheme: RIA; contract number: 645361). The authors acknowledge support from the technical staff of the James Watt Nanofabrication Centre at Glasgow University.